 \documentclass[twocolumn,showpacs,preprintnumbers,amsmath,amssymb]{revtex4}

 \usepackage{amsmath} 
\usepackage{amsfonts} 
\usepackage{amssymb}
\usepackage{graphicx} 
\makeatletter
\DeclareRobustCommand\MakeTextUppercase{%
  \@uclcnotmath{\def\i{I}\def\j{J}}{##1##2}\large}
\DeclareRobustCommand\MakeTextLowercase{%
  \@uclcnotmath{}{##2##1}\lowercase}
\protected@edef\MakeTextLowercase#1{\MakeTextLowercase{#1}}
\let\ProvidesPackage\ProvidesPackage@latex
\let\ProcessOptions\ProcessOptions@latex
\let\DeclareOption\DeclareOption@latex
\expandafter
\let\csname MakeUppercase \expandafter\endcsname
    \csname MakeTextUppercase \endcsname
\expandafter
\let\csname MakeLowercase \expandafter\endcsname
    \csname MakeTextLowercase \endcsname
\appdef\class@documenthook{%
 \switch@longtable
}%
\makeatother

\makeatletter
\relax
  \def\ps@headings{%
      \let\@oddfoot\@empty\let\@evenfoot\@empty
      \def\@evenhead{\thepage\hfil\slshape\leftmark}%
      \def\@oddhead{{\slshape\rightmark}\hfil\thepage}%
      \let\@mkboth\markboth
    \def\sectionmark##1{%
      \markboth {%
        \ifnum \c@secnumdepth >\z@
          \thesection\quad
        \fi
        ##1}{}}%
    \def\subsectionmark##1{%
      \markright {%
        \ifnum \c@secnumdepth >\@ne
          \thesubsection\quad
        \fi
        ##1}}}%
\def\ps@myheadings{%
    \let\@oddfoot\@empty\let\@evenfoot\@empty
    \def\@evenhead{\thepage\hfil\slshape\leftmark}%
    \def\@oddhead{{\slshape\rightmark}\hfil\thepage}%
    \let\@mkboth\@gobbletwo
    \let\sectionmark\@gobble
    \let\subsectionmark\@gobble
    }%
\def\ps@article{%
    \@provide\@evenhead{\let\\\heading@cr\thepage\quad\checkindate\hfil{\leftmark}}%
    \@provide\@oddhead{\let\\\heading@cr{\rightmark}\hfil\checkindate\quad\thepage}%
    \@provide\@oddfoot{}%
    \@provide\@evenfoot{}%
    \let\@mkboth\markboth
  \let\sectionmark\@gobble
  \let\subsectionmark\@gobble
}%
\def\ps@article@final{%
    \@provide\@evenhead{\let\\\heading@cr\thepage\quad\checkindate\hfil{\leftmark}}%
    \@provide\@oddhead{\let\\\heading@cr{\rightmark}\hfil\checkindate\quad\thepage}%
    \@provide\@oddfoot{}%
    \@provide\@evenfoot{}%
    \let\@mkboth\markboth
    \def\sectionmark##1{%
      \markboth{%
        \@ifnum{\c@secnumdepth >\z@}{\thesection\hskip 1em\relax}{}%
         ##1%
       }{}%
    }%
    \def\subsectionmark##1{%
      \markright {%
        \@ifnum{\c@secnumdepth >\@ne}{\thesubsection\hskip 1em\relax}{}%
         ##1%
      }%
    }%
}%
\makeatother

\makeatletter
\def\bibliographystyle{\def\@bibstyle}%
\def\bibsection{%
  \let\@hangfroms@section\@hang@froms
  \section*{\large\refname}%
  \@nobreaktrue
}%
\makeatother

\renewcommand\refname{References}

\begin{document}

\preprint{MIT-CTP  / 4348}
\title{Quantum Time Crystals}
\author{Frank Wilczek}
\vspace*{.2in}
\affiliation{Center for Theoretical Physics \\
Department of Physics, Massachusetts Institute of Technology\\
Cambridge Massachusetts 02139 USA}
\vspace*{.3in}

\begin{abstract}
Some subtleties and apparent difficulties associated with the notion of spontaneous breaking of time translation symmetry in quantum mechanics are identified and resolved.   A model exhibiting that phenomenon is displayed.   The possibility and significance of breaking of imaginary time translation symmetry is discussed.   
\end{abstract}
\maketitle

\bigskip


Symmetry and its spontaneous breaking is a central theme in modern physics.   Perhaps no symmetry is more fundamental than time translation symmetry, since time translation symmetry underlies both the reproducibility of experience and, within the standard dynamical frameworks, the conservation of energy.    So it is natural to consider the question, whether time translation symmetry might be spontaneously broken in a closed quantum-mechanical system.   That is the question we will consider, and answer affirmatively, here.   

Here we are considering the possibility of time crystals, analogous to ordinary crystals in space.  They represent spontaneous emergence of a clock within a time-invariant dynamical system.   Classical time crystals are considered in a companion paper \cite{shapereWilczek}; here the primary emphasis is on quantum theory. 

Several considerations might seem to make the possibility of quantum time crystals implausible.   The Heisenberg equation of motion for an operator with no intrinsic time dependence reads
\begin{equation}\label{noGo}
\langle \Psi | \dot {\cal O} | \Psi \rangle  ~=~ i \langle \Psi | [ H, {\cal O} ] | \Psi \rangle ~\rightarrow_{\Psi = \Psi_E}  ~ \ 0,
\end{equation}
where the last step applies to any eigenstate $\Psi_E$ of $H$.   This seems to preclude the possibility of an order parameter that could indicate the spontaneous breaking of infinitesimal time translation symmetry.   Also, the very concept of ``ground state'' implies state of lowest energy; but in any state of definite energy (it seems) the Hamiltonian must act trivially.   Finally, a system with spontaneous breaking of time translation symmetry in its ground state must have some sort of motion in its ground state, and is therefore perilously close to fitting the definition of a perpetual motion machine.

{\it Ring Particle Model}:  And yet there is a familiar physical phenomenon that almost does the job.  A superconductor, in the right circumstances, can support a stable current-carrying ground state.   Specifically, this occurs if we have a superconducting ring threaded by a flux that is a fraction of the flux quantum.   If the current is constant then nothing changes in time, so time translation symmetry is not broken; but clearly there is a sense in which something is moving.    

We can display the essence of this situation in a simple model, that displays its formal structure clearly.   Consider a particle with charge $q$ and unit mass, confined to a ring of unit radius that is threaded by flux $2 \pi \alpha/q$.    The Lagrangian, canonical (angular) momentum, and Hamiltonian for this system are respectively
\begin{eqnarray}
L ~&=&~ \frac{1}{2} \dot \phi^2 + \alpha \dot \phi, \nonumber \\
\pi_\phi ~&=&~ \dot \phi + \alpha, \nonumber \\
H ~&=&~ \frac{1}{2} (\pi_\phi - \alpha)^2
\end{eqnarray}
$\pi_\phi$, through its role as generator of (angular) translations, and in view of the Heisenberg commutation relations, is realized as $-i\frac{\partial}{\partial \phi}$.   Its eigenvalues are integers $l$, associated with the states $| l \rangle = e^{il\phi}$.   For these states we have
\begin{eqnarray}
\langle l | \dot \phi | l \rangle ~&=&~ l - \alpha, \nonumber \\
\langle l | H | l \rangle ~&=&~ \frac{1}{2}(l - \alpha)^2 .
\end{eqnarray}
The lowest energy state will occur for the integer $l_0$ that makes $l-\alpha$ smallest.   If $\alpha$ is not an integer, we will have 
\begin{equation}\label{dotPhi}
\langle l_0 | \dot \phi | l_0 \rangle ~=~ l_0 - \alpha ~\neq~ 0 .
\end{equation}

The case when $\alpha$ is half an odd integer requires special consideration.  In that case we will have two distinct states $|\alpha \pm \frac{1}{2} \rangle$ with the minimum energy.   We can clarify the meaning of that degeneracy by combining two simple observations.  First, that the combined operation $G_k$ of multiplying wave functions by $e^{i k \phi}$ and changing $\alpha \rightarrow \alpha +k$, for integer $k$, in the Lagrangian leaves the dynamics invariant.  Indeed, if we interpret $\alpha$ in $L$ as embodying a constant gauge potential,  $G_k$ is a topologically non-trivial gauge transformation on the ring, corresponding to the multiply-valued gauge function $A\rightarrow  A +  \nabla \Lambda$, $\Lambda = k\phi/q$.   Note that the total flux is {\it not\/} invariant under this topologically non-trivial gauge transformation, which cannot be extended smoothly off the ring, so $L$ is modified.   Second, that the operation of time-reversal $T$, implemented by complex conjugation of wave-functions, takes $|l\rangle \rightarrow |-l\rangle$ and leaves the dynamics invariant if simultaneously $\alpha \rightarrow -\alpha$.   Putting these observations together, we see that the combined operation
\begin{equation}\label{modifiedT}
\tilde T ~=~ G_{2\alpha} T  
\end{equation}
leaves the Lagrangian invariant; it  is a symmetry of the dynamics and maps $| l \rangle \rightarrow | 2\alpha -l\rangle$.  $\tilde T$ interchanges $|l \pm \alpha \rangle \rightarrow |l \mp \alpha \rangle$.   Thus the degeneracy between those states is a consequence of a modified time-reversal symmetry.    We can choose combinations $| \alpha + \frac{1}{2} \rangle \pm | \alpha - \frac{1}{2} \rangle$ that simultaneously diagonalize $H$ and $\tilde T$; for these combinations the expectation value of $\dot \phi$ vanishes.     

Returning to the generic case: For $\alpha$ that are not half-integral time-reversal symmetry is not merely modified, but simply broken, and there is no degeneracy.   How do we reconcile $\langle l_0 | \dot \phi | l_0 \rangle ~\neq~ 0$ with  Eqn. (\ref{noGo})?  The point is that $\dot \phi$, despite appearances, is neither the time-derivative of a legitimate operator nor the commutator of the Hamiltonian with one, since $\phi$, acting on wave functions in Hilbert space, is multivalued.    By way of contrast operators corresponding to single-valued functions of $\phi$, spanned by trigonometric functions ${\cal O}_k = e^{ik\phi}$, do satisfy Eqn. (\ref{noGo}) for the eigenstates $|\Psi \rangle = | l\rangle$.   

Wave functions of the quantized ring particle model correspond to the (classical) wave functions that appear in the Landau-Ginzburg theory of superconductivity.  Those wave functions, in turn, heuristically describe the wave function for macroscopic occupation of the single-particle quantum state appropriate to a Cooper pair, regarded as a particle.   Under this correspondence, the non-vanishing expectation value of $\dot \phi$ for the ground state of the ring particle subject to fractional flux maps onto the persistent current in a superconducting ring.

{\it Symmetry Breaking and Observability}: As mentioned previously, the choice of a ground state that violates time translation symmetry $\tau$ must be based on some criterion other than energy minimization.   But what might seem to be a special difficulty with breaking $\tau$, because of its connection to the Hamiltonian, actually arises in only slightly different form for all cases of spontaneous symmetry breaking.  Consider for example the breaking of number (or dually, phase) symmetry.   We characterize such breaking through a complex order parameter, $\Phi$, that acquires a non-zero expectation value, which we can take to be real:
\begin{equation}
\langle 0  | \Phi | 0 \rangle ~=~ v ~\neq 0 .
\end{equation}

We also have states $| \sigma \rangle$ related to $| 0 \rangle$ by the symmetry operation.  These are all energetically degenerate and mutually orthogonal (see below), and satisfy 
\begin{equation}
\langle \sigma  | \Phi | \sigma \rangle ~=~ v e^{i\sigma}.
\end{equation}
The superposition 
\begin{equation}
| \Omega \rangle ~=~ \frac{1}{2\pi} \int\limits^{2\pi}_0 \, d\sigma \ | \sigma \rangle
\end{equation}
is energetically degenerate with all the $|\sigma \rangle$, and it is symmetric, with
\begin{equation} 
\langle \Omega | \Phi | \Omega \rangle ~=~ 0.
\end{equation}

Why then do we prefer one of the states $| \sigma \rangle$ as a description of the physical situation?  The reason is closely related to the emergent orthogonality of the different $| \sigma \rangle$ states, as we now recall.   We envisage that our system extends over a large number $N$ of identical subsystems having correlated values of the long-range order parameter $\sigma$, but otherwise essentially uncorrelated.  Then we can express the total wave function in the form
\begin{equation}
\Psi_\sigma (x_1, ... , x_N) ~\approx~ \prod\limits_{j=1}^N \psi_\sigma (x_j).
\end{equation}
For different values $\sigma, \sigma^\prime$ we have therefore
\begin{equation}
\langle \Psi_{\sigma^\prime} | \Psi_\sigma \rangle \approx \prod\limits_{j=1}^N \langle \psi_{\sigma^\prime}( x_j) | \psi_\sigma ( x_j) \rangle 
=( f_{\sigma^\prime \sigma})^N \rightarrow 0 
\end{equation}
for  $\sigma^\prime \neq \sigma$ and large $N$, since $| f_{\sigma^\prime \sigma} | < 1$.   Similarly, for any finite set of local observables (that is, observables whose arguments include only upon a finite subset of the $x_j$), we have
\begin{equation}\label{observables}
\langle \Psi_{\sigma^\prime} | {\cal O}_1 (x_a){\cal O}_2 (x_b) ... | \Psi_\sigma \rangle \propto (f_{\sigma^\prime \sigma})^{N-{\rm finite}} \rightarrow  0
\end{equation}
for $\sigma^\prime \neq \sigma$.  Since the off-diagonal matrix elements vanish, any world of local observations (including ``observations'' by the environment) can be described using a single $|\sigma \rangle$ state.  Averaging over them, to produce $| \Omega \rangle$, is a purely formal operation.   Measurement of a non-singlet observable will project onto a $| \sigma \rangle$ state.   


This analysis \cite{strocchi} brings out several relevant points.  The physical criterion that identifies useful ``ground states'' is not simply energy, but also robust observability.   Mathematically, that requirement is reflected in the orthogonality of the Hilbert spaces built upon $| \sigma \rangle$ states by the action of physical observables.   The large $N$ limit is crucial for spontaneous symmetry breaking.   It is only in that infinite degree-of-freedom, or (as it is usually called) infinite volume, limit, that the $| j \rangle$ and $| \sigma \rangle$ states become degenerate, and the latter are preferred.   Important for present purposes: The preceding discussion applies, with only symbolic changes, when we consider possible breaking of time-translation $\tau$ in place of phase symmetry.

{\it Soliton Model}:  After these preparations, it is not difficult to construct an appropriate model.   We consider a large number of ring-particles with an attractive interaction.   Heuristically, we can expect that they will want to form a lump and, in view of Eqn. (\ref{dotPhi}), that they will want to move.  So we can expect that the physical ground state features a moving lump, which manifestly breaks $\tau$.   

To make contact with the argument of the previous section, we need an appropriate notion of locality. For simplicity we assume that the particles have an additional integer label, besides the common angle $\phi$, and that the physical observables are of finite range in the additional label.   (Imagine an array of separate rings, displaced along an axis, so that the coordinates of particle $j$ are $(\phi, x = ja)$.)   I will return to these conceptual issues below, after describing the construction.  

An appropriate Hamiltonian is
\begin{eqnarray}
H ~ &=&~ \sum\limits_{j=1}^N \frac{1}{2} (\pi_j - \alpha)^2 - \frac{\lambda}{N-1} \sum\limits_{j \neq k, 1}^N \delta (\phi_j - \phi_k) ~\\ 
   &\equiv& ~ \sum\limits_{j=1}^N \frac{1}{2} (\pi_j - \alpha)^2 + V(\phi_1, ... , \phi_N), \nonumber
\end{eqnarray}
with the understanding that $H$ acts on periodic functions, so the $\delta$ interaction is well defined.   (Here the discrete index appears as a subscript.)   

We work in the mean field approximation, taking a product {\it ansatz\/}
\begin{equation}
\Psi (\phi_1, ... , \phi_n) ~=~ \prod\limits_{j=1}^N \psi (\phi_j),
\end{equation}
and solving an approximate one-body equation for $\psi$.     To get such an equation, we define an effective potential 
\begin{eqnarray}
V_{\rm eff.} (\phi_1, ... , \phi_N) ~&=&~ \sum\limits_{j=1}^N W(\phi_j) \nonumber \\ 
W (\phi_j) ~&=&~  \int \prod\limits_{k \neq j} \, d\phi_k \, \psi^*(\phi_k) V \psi(\phi_k),
\end{eqnarray}
so that 
\begin{equation}
\langle \Psi |  V_{\rm eff.} | \Psi \rangle ~=~ \langle \Psi | V | \Psi \rangle.
\end{equation}
Then the effective Schr\"odinger equation for $\Psi$, 
\begin{equation}
i\frac{\partial \Psi}{\partial t} ~=~ \bigl( \, \sum\limits_{j=1}^N \frac{1}{2} (\pi_j - \alpha)^2 + V_{\rm eff.} \bigr) \Psi ,
\end{equation} 
reduces to the one-body non-linear Schr\"odinger equation
\begin{equation}\label{nlS}
i\frac{\partial \psi}{\partial t} ~=~ \frac{1}{2}(\pi_\phi - \alpha)^2 \psi - \lambda | \psi |^2 \psi
\end{equation}
for $\psi$.

Consider first the case $\alpha = 0$.   Eqn. (\ref{nlS}) can be solved for a stationary state in terms of the Jacobi ${\rm dn}$ elliptic function, with 
\begin{eqnarray}
\psi(\phi, t) ~&=&~ e^{-i{\cal E}t} \psi_0 (\phi + \beta) \nonumber \\
\psi_0 (\phi ) ~&=&~ r\,  {\rm dn} \,  ( r \sqrt \lambda \phi, k^2) \nonumber \\
{\cal E} ~&=&~ - r^2 \lambda (1-\frac{k^2}{2} ) ,
\end{eqnarray}
with $\beta$ a disposable parameter.
To fix the parameters $k, r$ we must impose $2\pi$ periodicity in $\phi$ and normalize $\psi_0$.   Those conditions become
\begin{eqnarray}
E(k^2) ~&=&~ \frac{\sqrt \lambda}{2r} \nonumber \\
K(k^2) ~&=&~ \pi r \sqrt \lambda
\end{eqnarray}
in terms of the complete elliptic integrals $E(k^2), K(k^2)$.    We can solve $E(k^2)K(k^2) = \frac{\pi \lambda}{2}$ for $k^2$, given $\lambda$.  The minimum value of the left-hand side occurs at $k=0$ and corresponds to $\lambda = \frac{\pi}{2}$.   Here ${\rm dn}\, (u, 0)$ reduces to a constant, and ${\cal E} = -1/4$.      
As $\lambda$ increases beyond that value $k$ rapidly approaches 1, as does $E(k^2)$.    ${\rm dn} (u, k^2) \rightarrow {\rm sech} \,  u$ and ${\cal E} \rightarrow -\lambda^2/8$ in that limit.   Of course the constant solution with ${\cal E} = -\lambda/2\pi$ exists for any value of $\lambda$, but when $\lambda$ exceeds the critical value the inhomogeneous solution is more favorable energetically. 
These results have simple qualitative interpretations.  The hyperbolic secant is the famous soliton of the non-linear Schr\"odinger equation on a line.  If that soliton is not too big it can be deformed, without prohibitive energy cost, to fit on a unit circle.   The parameter $\beta$ reflects spontaneous breaking of (ordinary) translation symmetry.    Here that breaking is occurring through a kind of phase separation.    

Our Hamiltonian is closely related, formally, to the Lieb-Liniger model \cite{LL}, but because we consider ultra-weak ($\sim 1/N$) attraction instead of repulsion, the ground state physics is very different.  Since our extremely inhomogeneous approximate ground state does not support low-energy, long-wavelength modes (apart from overall translation), it has no serious infrared sensitivity.

Now since non-zero $\alpha$ can be interpreted as magnetic flux through the ring, we might anticipate, from Faraday's law, that as we turn it on, starting from $\alpha=0$,  our lump of charge will feel a simple torque.   (Note that since Faraday's law is a formal consequence of the mathematics of gauge potentials, its use does not require additional hypotheses.)    We can also apply ``gauge transformations'', as in the discussion around Eqn. (\ref{modifiedT}).   These observations are reflected mathematically in the following construction:  For any $l$, we solve 
\begin{equation}
i \frac{\partial \psi_l}{\partial t} ~=~ \frac{1}{2} (-i \partial_\phi - \alpha)^2 \psi_l - \lambda |\psi_l |^2 \psi_l,
\end{equation}
with
\begin{eqnarray}
\psi_l (\phi, t) ~&=&~ \, e^{-il\phi} \, \tilde \psi (\phi + (l +\alpha)t, t) \nonumber \\
i\frac{\partial \tilde \psi}{\partial t} ~&=&~ \frac{1}{2} (-i \partial_\phi)^2 \tilde \psi - \lambda | \tilde \psi |^2 \tilde \psi + \frac{(l+\alpha)^2}{2}\,  \tilde \psi .
\end{eqnarray}
As in the non-interacting ring particle model, the lowest energy is obtained by minimizing $l_0 + \alpha$, for integral $l_0$.   If $\alpha$ is not an integer $\psi_{l_0} (\phi, t)$ will be a moving lump, and time translation symmetry will have been spontaneously broken.   If $\alpha$ is half an odd integer, then its $\tilde T$ symmetry is spontaneously broken too.    

This example exhibits several characteristic features of natural $\tau$ breaking \cite{shapereWilczek}. The  lump moves along a constant energy trajectory.   The parameter $\beta$, which parameterizes an orbit of (ordinary) translation symmetry, changes at a constant rate; both $\tau$ and translation symmetry are broken, but a combination remains intact.   


Now let us return to address the conceptual issues alluded to earlier.   Our model Hamiltonian was non-local , but we required observables to be local.   That schizophrenic distinction can be appropriate, since the Hamiltonian might be -- and, for our rather artificial dynamics, would have to be -- carefully engineered, as opposed to being constructed from easily implemented, natural observables.    Moreover it is not unlikely that the assumption of all-to-all coupling could be relaxed, in particular by locating the rings at the nodes of a multidimensional lattice and limiting the couplings to a finite range.   

Were we literally considering charged particles confined to a common ring, and treating the electromagnetic field dynamically, our moving lump of charge would radiate.  The electromagnetic field provides modes that couple to all the particles, and in effect provide observers who manifestly violate the framework of Eqn. (\ref{observables}).   That permits, and enforces, relaxation to a $| k \rangle$ state.  Simple variations can ameliorate this issue, {\it e. g}.  use of multipoles in place of single charges, embedding the system in a cavity, or simply arranging that the motion is slow.   A more radical variation, that also addresses the unrealistic assumption of attraction among the charges, while still obtaining spatial non-uniformity, would be to consider charged particles on a ring that form -- through repulsion! -- a Wigner lattice.


{\it Imaginary Time Crystals}:  In the standard treatment of finite temperature quantum systems using path integral techniques, one considers configurations whose arguments involve imaginary values of the time, and imposes imaginary-time periodicity in the inverse temperature $\beta = 1/T$.    In this set-up the whole action is converted, in effect, into a potential energy: time derivatives map onto gradients in imaginary time, which is treated on the same footing as the spatial variables.  

At the level of the action, there is symmetry under translations in imaginary time (iTime).   But since iTime appears, in this formulation, on the same footing as the spatial variables, it is natural to consider the possibility that for appropriate systems the dominant configurations in the path integral are iTime crystals.    Let the iTime crystal have preferred period $\lambda$.  When $\beta$ is an integer multiple of $\lambda$ the crystal will fit without distortion, but otherwise it must be squeezed or stretched, or incorporate defects.    Periodic behavior of thermodynamics quantities in $1/T$, with period $\lambda$, arise, and provide an experimental diagnostic.   Integration over the collective coordinate for the broken symmetry contributes to the entropy, even at zero temperature.   Inspired by the spatial crystal - iTime crystal analogy, one might also consider the possibility of iTime glasses (iGlasses), which would likewise have residual entropy, but no simple order, or iQuasicrystals.


{\it Comments}:  1. It is interesting to speculate that a (considerably) more elaborate quantum-mechanical system, whose states could be interpreted as collections of qubits, might be engineered to traverse, in its ground configuration, a programmed landscape of structured states in Hilbert space over time.  

2. Fields or particles in the presence of a time crystal background will be subject to energy-changing processes, analogous to crystalline Umklapp processes.   In either case the apparent non-conservation is in reality a transfer to the background.  (In our earlier model, $O(1/N)$ corrections to the background motion arise.)

3. Many questions that arise in connection with any spontaneous ordering, including the nature of transitions into or out of the order at finite temperature, critical dimensionality, defects and solitons, and low-energy phenomenology, likewise pose themselves for time crystallization.   There are also interesting issues around the classification of space-time periodic orderings (roughly speaking, four dimensional crystals \cite{4crystals}).   

4. The a.c. Josephson effect is a semi-macroscopic oscillatory phenomenon related in spirit to time crystallization.   It requires, however, a voltage difference that must be sustained externally.  

5. Quantum time crystals based on the classical time crystals of \cite{shapereWilczek}, which use singular Hamiltonians, can be constructed by combining the ideas of this paper with those of \cite{henneaux}, \cite{branched}.   The appearance of swallowtail band structures in \cite{swallowtail}, and emergence of complicated frequency dependence in modeling finite response times \cite{shapereWilczek}, as in \cite{dmft}, suggest possible areas of application.   


\bigskip

{\it Acknowledgements} I thank B. Halperin, Hong Liu, J. Maldacena, and especially Al Shapere for helpful comments.   This work is supported in part by DOE grant DE-FG02-05ER41360.

\end{document}